%% file: main.tex
\documentclass[sigconf]{acmart}
\usepackage[T1]{fontenc}
\usepackage{graphicx}
\usepackage{xkeyval}
\usepackage{csquotes}
\usepackage{subfigure}
\usepackage{algorithm}
\usepackage{algpseudocode}
\usepackage{enumitem}
\usepackage{amsmath}
\usepackage{multirow}
\usepackage{tabularx}
\usepackage{array}
\usepackage{url}
\usepackage{fixltx2e}
\usepackage{color}
\usepackage{adjustbox}
\newcolumntype{P}[1]{>{\centering\arraybackslash}p{#1}}
\usepackage{wasysym}
\setcopyright{rightsretained}
\newtheorem{defn}{Definition}[section]

\begin{document}
\copyrightyear{2018}
\acmYear{2018} 
\setcopyright{iw3c2w3}
\acmConference[WWW 2018]{The 2018 Web Conference}{April 23--27, 2018}{Lyon, France}
\acmBooktitle{WWW 2018: The 2018 Web Conference, April 23--27, 2018, Lyon, France}
\acmPrice{}
\acmDOI{https://doi.org/10.1145/3178876.3186119}
\acmISBN{978-1-4503-5639-8}

\fancyhead{}

\title{Collective Classification of Spam Campaigners on Twitter: A Hierarchical Meta-Path Based Approach}

\author{Srishti Gupta}
\affiliation{%
  \institution{IIIT-Delhi}
  }
\email{srishtig@iiitd.ac.in}

\author{Abhinav Khattar}
\affiliation{%
  \institution{IIIT-Delhi}
  }
\email{abhinav15120@iiitd.ac.in}

\author{Arpit Gogia}
\affiliation{%
  \institution{DTU}
  }
\email{aarpitgogia@gmail.com}

\author{Ponnurangam Kumaraguru}
\affiliation{%
  \institution{IIIT-Delhi}
  }
\email{pk@iiitd.ac.in}

\author{Tanmoy Chakraborty}
\affiliation{%
  \institution{IIIT-Delhi}
  }
\email{tanmoy@iiitd.ac.in}

\begin{abstract}
Cybercriminals have leveraged the popularity of a large user base available on Online Social Networks~(OSNs) to spread spam campaigns by propagating phishing URLs, attaching malicious contents, etc. However, another kind of spam attacks using phone numbers has recently become prevalent on OSNs, where spammers advertise phone numbers to attract users' attention and convince them to make a call to these phone numbers. The dynamics of phone number based spam is different from URL-based spam due to an inherent trust associated with a phone number. While previous work has proposed strategies to mitigate URL-based spam attacks, phone number based spam attacks have received less attention. 

In this paper, we aim to detect spammers that use phone numbers to promote campaigns on Twitter. To this end, we collected information (tweets, user meta-data, etc.) about $3,370$ campaigns spread by $670,251$ users. We model the Twitter dataset as a {\em heterogeneous network} by leveraging various interconnections between different types of nodes present in the dataset.
In particular, we make the following contributions -- (i) We propose a simple yet effective metric, called {\em Hierarchical Meta-Path Score}~({\em HMPS}) to measure the proximity of an unknown user to the other known pool of spammers. (ii) We design a {\em feedback-based active learning strategy} and show that it significantly outperforms three state-of-the-art baselines for the task of spam detection. Our method achieves 6.9\% and 67.3\% higher F1-score
and AUC, respectively compared to the best baseline method. (iii) To overcome the problem of less training instances for supervised learning, we show that our proposed {\em feedback strategy} achieves 25.6\% and 46\% higher F1-score and AUC respectively than other oversampling strategies. Finally, we perform a case study to show how our method is capable of detecting those users as spammers who have not been suspended by Twitter (and other baselines) yet.
\end{abstract}


  
\maketitle
\input{intro}
\input{problem}
\input{method}
\input{results}
\input{related}
\input{conclusion}

\bibliographystyle{ACM-Reference-Format}
\bibliography{ref} 
\end{document}

%% file: intro.tex
\section{Introduction}
Online Social Networks (OSNs) are becoming more and more popular in the recent years, used by millions of users. As a result, OSNs are being abused by spam campaigners to carry out phishing and spam attacks~\cite{grier2010spam}. While attacks carried using URLs~\cite{grier2010spam, gao2010detecting,chu2012detecting,zhang2016detecting,thomas2011suspended} has been extensively explored in the literature, attacks via a new action token, i.e., a {\em phone number} is mostly unexplored. 
Traditionally, spammers have been exploiting telephony system in carrying out social engineering attacks either by calling victims or sending SMS~\cite{yeboah2014phishing}. Recently, spammers have started abusing OSNs where they float phone numbers controlled by them. 
Besides exploiting trust associated with a phone number, spammers save efforts in reaching out their victims themselves. 

\begin{figure}[h]
\includegraphics[width=0.6\linewidth]{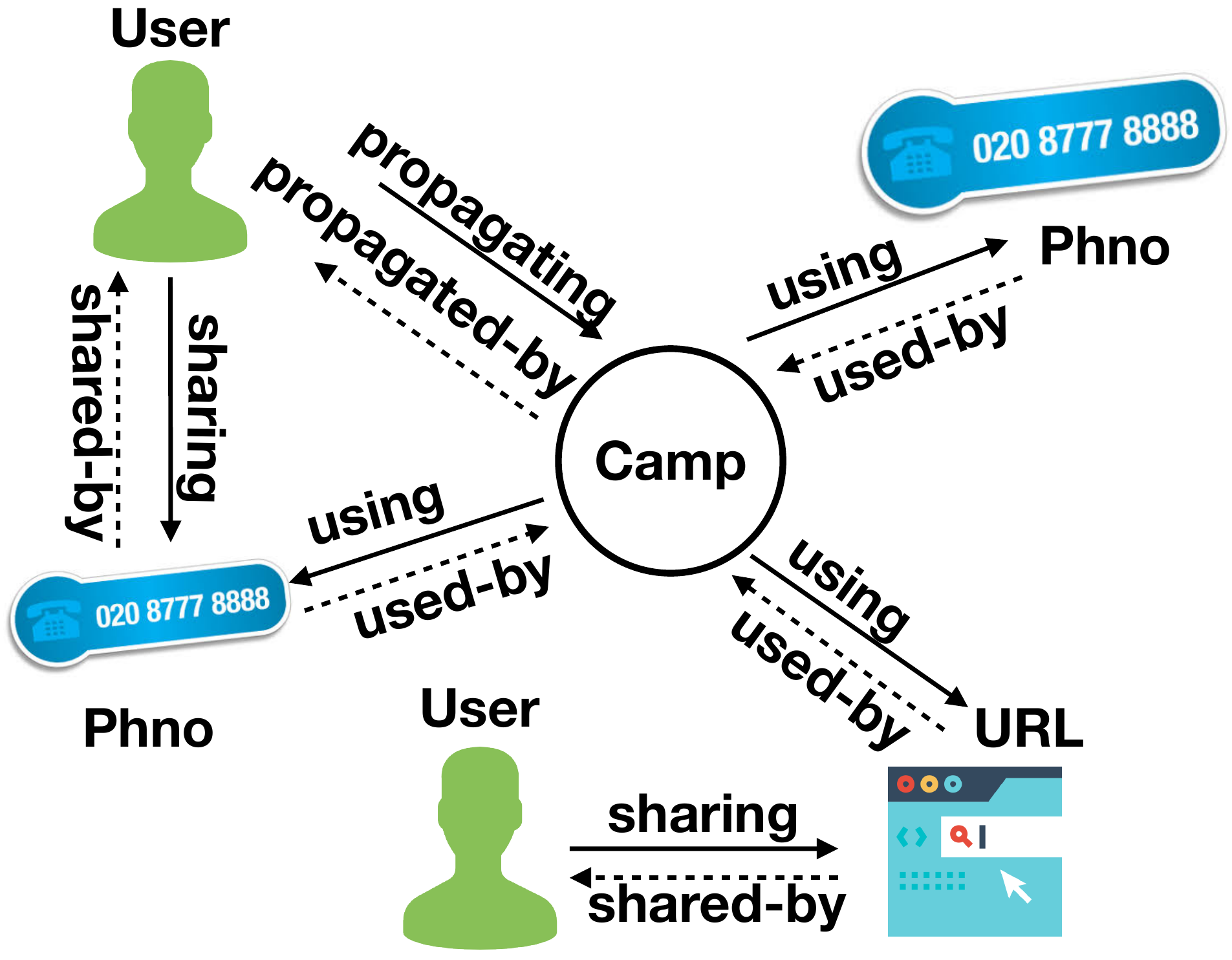}
\caption{Twitter modeled as a heterogeneous network.}
\label{fig:hetero}
\vspace{-4mm}
\end{figure}

\textbf{Present Work: Problem definition}. In this paper, we aim to detect spam campaigners ({\em aka}, spammers) spreading spam campaigns using phone numbers on Twitter. We here define \emph{spammers} as user accounts that use phone numbers to aggressively promote products, disseminate pornography, entice victims for lotteries and discounts, or simply mislead victims by making false promises. 
Discovering the correspondence between the spammer accounts and the resources (such as URL or phone number) used for spam activities is a crucial task. As the phone numbers are being propagated by spammers, and their monetization revenue starts once people call them, it is fair to assume that these phone numbers would be under their control. As an added advantage of this, if we can identify the spammer accounts in Twitter and bring them down, the entire campaign would get disintegrated. 
To identify spammers, we model the Twitter dataset as a heterogeneous graph where there are different connections between heterogeneous type of entities: users, campaigns, and the action tokens (phone number or URL) as shown in Figure~\ref{fig:hetero}.  
Heterogeneous networks have been proposed for data representation in a variety of datasets like path similarity in scholar data~\cite{sun2011pathsim}, link prediction in social network data~\cite{kong2013inferring}, etc. Objects of different types and links carry different semantics. For instance, a phone number being a more stable resource would help in connecting user accounts over an extended period. Physical identity verification is required to purchase phone numbers, while only e-mail verification is sufficient to purchase domains. Studying similarity between a pair of nodes keeping the heterogeneous nature of the network helps in distinguishing the semantics of different types of paths connecting these two nodes. 
To distinguish the semantics among paths connecting two nodes,
we introduce a \emph{meta-path}  based similarity framework for nodes
of the same type in the heterogeneous network. A meta-path
is a sequence of relations between node types, which defines a
new composite relation between its starting type and ending type.
It provides a powerful mechanism to classify objects sharing similar semantics appropriately.

\textbf{Present Work: Motivation of the work.} The problem of identifying spammers on Twitter that use phone numbers is useful in many aspects. Attacks using phone numbers and URLs are different in some aspects: in URL based spam, the campaign propagates and spreads on the same medium, i.e., OSNs, while in case of phone number based spam, the attacking medium is a telephone and the propagating medium is OSNs. As a result, it is challenging for OSN service providers to track down the accounts spreading these spam campaigns. In addition, there is no meta-data available for phone numbers, unlike URLs where landing page information, length of URLs, obfuscation, etc. can be checked. Perhaps due to the challenges associated with finding spam phone numbers, there have been several attacks and financial losses caused by the phone-based attacks~\cite{miramirkhani2017dial}. Using the collective classification approach proposed in the paper, Twitter will be able to find potential spammers and suspend the accounts, thereby restricting phone number based spam campaigns.

\textbf{Present Work: A collective classification approach for detecting spam campaigners.} In this work, we use the \emph{collective classification} approach that exploits the dependencies of a group of linked nodes where some class labels are known and labels diffuse to other unknown nodes in the network. In our case, the known nodes are the already suspended Twitter users that were propagating campaigns with phone numbers.  
Here, we propose {\em Hierarchical Meta-Path Score}~({\em HMPS}), a simple yet effective similarity measure between a pair of nodes in the heterogeneous network. We first build campaign-specific hierarchical trees from the large heterogeneous network. We then systematically extract most relevant meta-paths from a pool of meta-paths linking various heterogeneous nodes. 

We collected tweets and other meta-data information of users from April-October, 2016, and identified $3,370$ campaigns, containing $670,251$ users (Section~\ref{dataset}). Each tweet carries a phone number. We consider user accounts suspended by Twitter as ground-truth spammers. However, due to the lack of enough training samples per campaign, we introduce a novel {\em feedback-based active learning mechanism} that uses a SVM-based one-class classifier for each campaign. Over multiple iterations, it keeps accumulating evidences from different campaigns to enrich the training set for each campaign. This, in turn, enhances the prediction performance of individual classifiers. The process terminates when there is no chance of finding the label of the unknown users across iterations.

\textbf{Summary of the evaluation.} We compare our model with three state-of-the-art baselines used for spam detection (Section~\ref{baseline}). We design various experimental setup to perform a thorough evaluation of our proposed method. 
We observe that our model outperforms the best baseline method by achieving 44.8\%, 16.7\%, 6.9\%, 67.3\% higher performance in terms of accuracy, precision, F1-score and AUC~(Section~\ref{baseline}).
We further demonstrate how~/~why one-class classifier~(Section~\ref{oneclasstwoclass}), active learning~(Section~\ref{feedbackresults}) and feedback-based learning~(Section~\ref{feedbackoversample1}) are better than 2-class classifier, general learning and other oversampling method, respectively. Moreover, we conduct a case study and present an intuitive justification why our method is superior to the other methods (Section~\ref{hmpsgood}). 

%% file: problem.tex
\section{Dataset} \label{dataset}
We collected tweets containing phone numbers from Twitter based on an exhaustive list of 400 keywords via Twitter streaming API. 
We chose Twitter due to easy availability of data. The data was collected from April - October, 2016. Since we intended to detect campaigns around phone numbers, the keywords we chose were specific to phone number such as `call', `SMS', `WA', `ring' etc. We accumulated $\sim22$ million tweets, each of which containing at least one phone number. The reason behind collecting only tweets containing phone numbers is that they are found to be a stable resource, i.e., spammers use them for a long period due to attached cost. Moreover, the phone numbers are known to help in forming better user communities~\cite{costin2013role}, which is the basis of the approach adopted in this work.

\textbf{Campaign identification:}  \label{campaignidentification}
We define a \emph{campaign} as a group of similar posts shared by a set of users propagating multiple phone numbers. A phone number could be a part of multiple campaigns; however, in this work, we restrict the phone number to be part of a single campaign (since our campaign detection approach is text-based, we want the campaigns to be coherent). Note that, multiple phone numbers could be a part of a single campaign. The detailed approach for campaign identification is shown in Figure~\ref{fig:clustering} using a toy example for three phone numbers as described below:
\begin{figure}[h]
\includegraphics[width=\columnwidth]{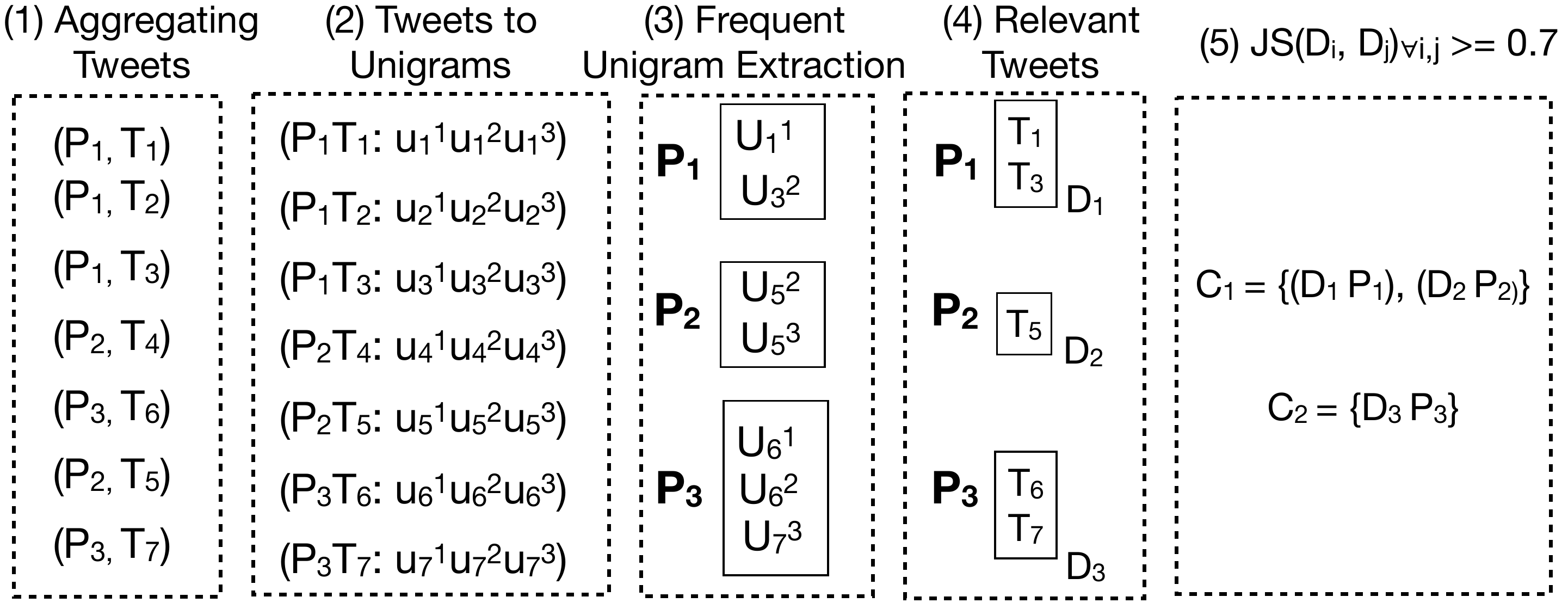} 
\caption{A schematic diagram of the framework for campaign identification (notation: $P$: a phone number, $U$: a unigram, $T$: a tweet represented by a set of unigrams, $D$: a document consisting of a set of similar tweets, and $C$: a campaign containing a document and its associated phone number).} 
\label{fig:clustering}
\end{figure}

\textbf{Step 1: Aggregating tweets.} For every phone number, we aggregate all the tweets containing that phone number in a set. We do not find a single tweet containing two phone numbers in our dataset. This implies that every phone number $P_{i}$ has a set of unique tweets represented as {$T_{1}, T_{2}, T_{3}, \cdots$}. In Figure \ref{fig:clustering}, $P_1$ is associated with $\{T_1,T_2,T_3\}$.

\textbf{Step 2: Tweets to unigrams.} We extract unigrams from tweets. Each tweet $T_i$ is now represented as $\{U_i^1, U_i^2, U_i^3, \cdots\}$. In Figure \ref{fig:clustering}, $T_1$ is represented as $\{U_1^1,U_1^2,U_1^3\}$.

\textbf{Step 3: Extracting frequent unigram.} We aggregate all tweets containing a certain phone number and extract top 30 unigrams that frequently appear in these tweets. This set of unigrams characterizes the document associated with the phone number. In Figure \ref{fig:clustering}, the set $\{U_1^1, U_3^2\}$ represents the document associated with $P_1$.

\textbf{Step 4: Selecting relevant tweets.} From the set of tweets associated with a certain phone number, we choose those which have at least 5 unigrams common with the set of 30 unigrams representing the phone number. In Figure \ref{fig:clustering}, we only choose $T_1$ and $T_3$ to form document $D_1$ for $P_1$  (note, in this example we only match at least one unigram in each tweet instead of 5 to be qualify as a part of the document). 

\textbf{Step 5: Jaccard similarity to find campaigns.} Once we form the document corresponding to a phone number, we use Jaccard coefficient to find similarity between two documents and combine them as part of the same campaign. If the Jaccard coefficient is greater than $0.7$ (experimentally calculated, as corresponding Silhouette score is $0.8$), the documents are merged and thus the corresponding phone numbers become part of a single campaign. In Figure \ref{fig:clustering}, $D_1$ and $D_2$ are merged together and form campaign $C_1$.

\begin{figure}[h]
\begin{center}
\subfigure[Spanish campaign sending porn videos on WhatsApp.]
{ \label{fig:spanish}
\includegraphics[width=0.45\linewidth]{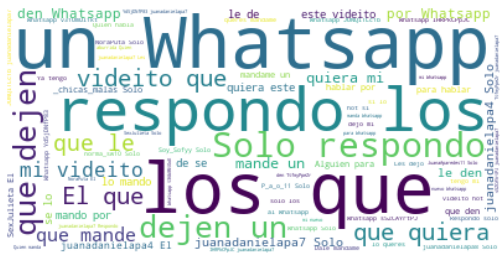}
}
\subfigure[Party reservation campaign offering discounted rate tickets.]
{ \label{fig:allcamp_vis}
\includegraphics[width=0.45\linewidth]{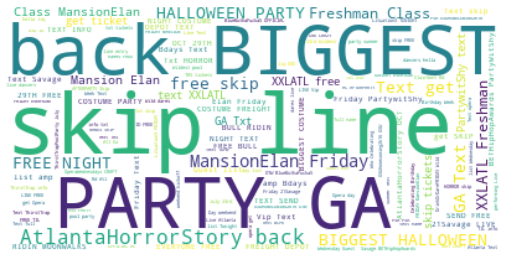}
}
\vspace{-3mm}
\caption{Word cloud of top two campaigns containing maximum suspended users.}
\label{fig:wordcloud}
\vspace{-5mm}
\end{center}
\end{figure}

Using this approach, we identify $22,390$ campaigns in our dataset. These account for $10,962,350$ tweets posted by $670,257$ users, containing $26,610$ unique phone numbers, and $893,808$ URLs. 
For collective classification to identify campaign-specific spammers, we need to have a set of labeled users. Therefore, we check the user accounts already suspended by Twitter. This process consists of a bulk query to Twitter's API with the profile ID of the account. Twitter redirects to \url{http://twitter.com/suspended}, and returns`error 404' in case the user account is suspended. Since Twitter suspension algorithm can have a delay in suspension, we made this query 6 months after the data collection. We find $5,593$ user accounts to be already suspended by Twitter. These accounts are taken later as the training set to perform spam classification (see Section \ref{sec:method}). Note that for further analysis, we take campaigns that have at least one suspended user --  $3,370$ out of $22,390$ campaigns ($670,251$ user accounts) are seen to observe such behavior. We also observe 21\% users to be part of multiple campaigns  (see Figure~\ref{fig:allaboutusers}(b)).
Figure~\ref{fig:wordcloud} shows the word cloud of top 2 campaigns containing the highest number of suspended users -- first one is a Spanish campaign~(Figure~\ref{fig:wordcloud}(a)) where people requested others to send WhatsApp invitation for receiving adult and pornographic videos. The second campaign~(Figure~\ref{fig:wordcloud}(b)) offers people reservations for parties and clubs at discounted rates. 

\section{Heterogeneous Information Networks (HIN)} \label{HIN}
We model the entire Twitter data as a heterogeneous information network (HIN).
A HIN is a special type of information
network which contains multiple types of users or links between users~\cite{sun2011pathsim}. Mathematically, it is defined as follows:
\begin{defn}
{\bf Heterogeneous network.}
A large number of OSNs are heterogeneous in nature, involving diverse relationship between the nodes. Heterogeneous network is represented as a graph, $G = \{V, E, T\}$ in which each node $v\in V$ and each link $e\in E$ are associated with their mapping functions: 
\[ \phi(v): V \rightarrow T_{V}\] \[\phi(e) : E \rightarrow T_{E}\]
respectively. $T_{V}\in T$ and $T_{E}\in T$ denote the sets of users and edge types.
\end{defn}

Our heterogeneous network contains different types of nodes such as users, campaign, URLs, and phone numbers; edges connecting two nodes represent different types of relationships (see Figure~\ref{fig:hetero}). A user is linked to a campaign by \texttt{promoting} or \texttt{promoted-by} relation; a user is linked to a phone-number by \texttt{sharing} or \texttt{shared-by} relation;  a campaign is linked to a phone-number by \texttt{using} or {\tt used-by} relations. 
Two users can be connected via different paths viz. {\tt user-phone-user}, {\tt user-url-user}, {\tt user-phone-url-user}, and so on.  Formally, these paths are called {\em meta-paths}, as defined below.
\begin{defn}
{\bf Meta-path.} A meta-path $\Pi_{1.....k}$ is a path defined on
the graph of network schema $T_{G}$ = ($U, R$), and is denoted in the
form of $U_{1} \xrightarrow{R_{1}} U_{2} \xrightarrow{R_{2}} U_{3} ..... \xrightarrow{R_{k}} U_{k+1}$ which defines a composite relation $R = R_{1} \circ R_{2} \circ  .... \circ  R_{l}$ between type $U_{1}$ and $U_{k+1}$, where $\circ$ denotes the composition operator on relations. In our context, $U~\in$~\{user, campaign, phone number, URL\} and $R~\in$~\{sharing, promoting, using\}.
\end{defn}

\begin{figure}[h]
\begin{center}
\includegraphics[width=0.6\linewidth]{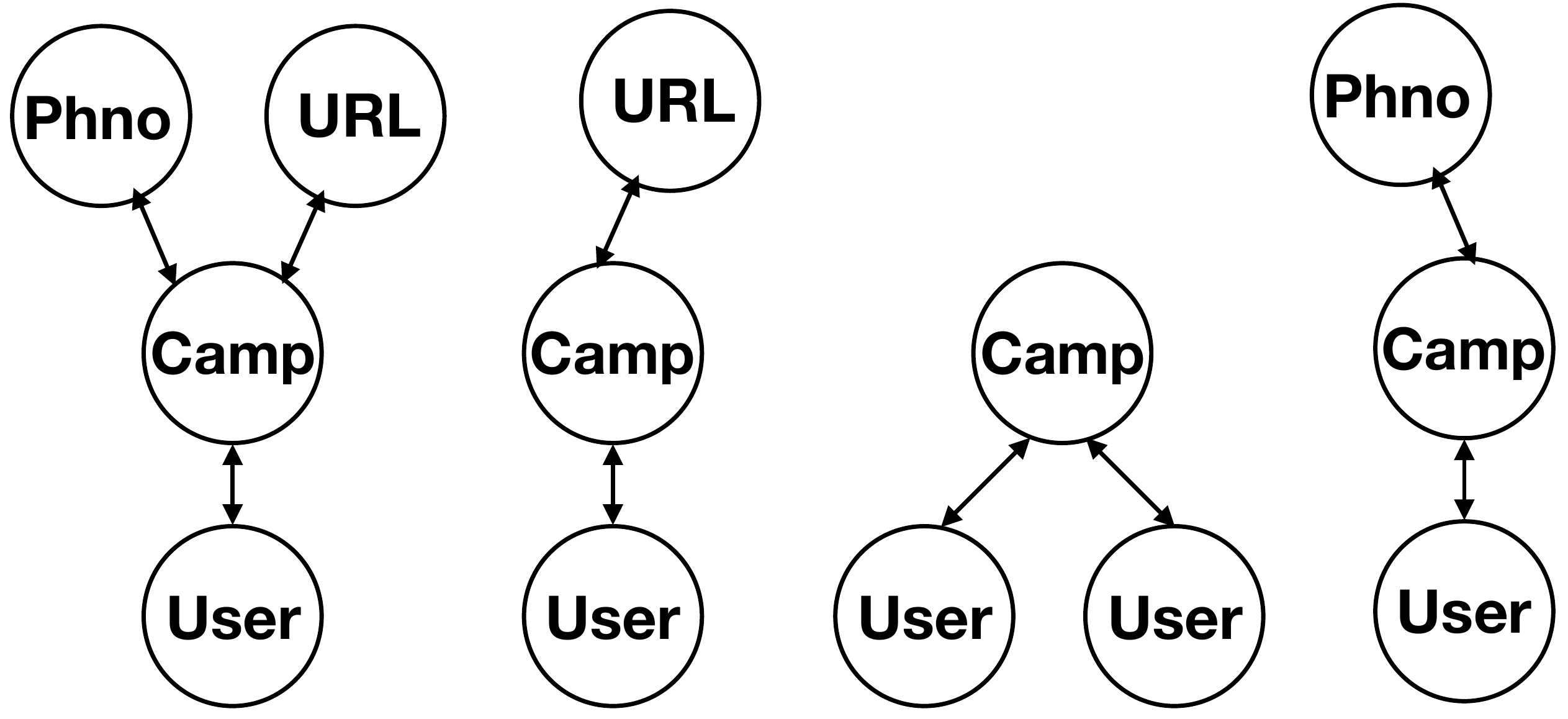}
\caption{Examples of different meta-paths present in the network.}
\label{fig:metapath}
\end{center}
\end{figure}

The length of a meta-path $\Pi$ is the number of relations that exist in $\Pi$ -- e.g., \texttt{user-phone-user} is a 2-length meta-path between a pair of users, while a 3-length meta-path instance between two users is
{\tt user-phone-URL-user}. Figure~\ref{fig:metapath} depicts some example meta-paths in our heterogeneous network.
For instance, a user participating in a campaign and sharing a phone number can be represented by a 2-length meta-path {\tt User-Camp-Phno}.

Given a user-specific meta-path, $\Pi$~=~{${U_{1}, U_{2}},.......U_{t}$}, similarity measures can be defined for a pair of users $x \in U_{1}$ and $y\in U_{2}$ according to the path instances between them following the meta-path.
Previous research has shown that including redundant meta-paths (i.e., a smaller meta-path that can be a part of a longer meta-path) in the collective classification may inject noise in the feature space, which can lead to over-fitting~\cite{kong2012meta}. To minimize the risk, it is advisable to extract meta-paths that cannot be further disintegrated to shorter meta-paths. The major challenge in dealing with meta-paths is to find {\em all and only relevant} meta-paths. Sun et al.~\cite{sun2011pathsim} showed that finding all possible meta-paths and picking the most relevant out of them is an NP-hard problem, and therefore many greedy approaches have been proposed to find relevant meta-paths~\cite{meng2015discovering}.
To the best of our knowledge, {\em this is the first work towards modeling Twitter as a heterogeneous network for spam campaigner detection by extracting relevant meta-paths}. Therefore, there is no prior work suggesting possible and relevant meta-paths for our heterogeneous network. 
To deal with these challenges, we propose a simple yet efficient concept, called \textbf{\textit{Hierarchical Meta-Path Scores}}~(\textbf{\textit{HMPS}}) to find similarity between a pair of users by picking shortest and relevant meta-paths (restricted to length  $4$) which can be used to calculate similarity between nodes.~\footnote{We experimented with meta-paths of length more than $4$. The results were not that encouraging compared to the time it takes to extract long-length meta-paths.} We also impose an additional constraint on the meta-path selection - we only consider intermediate nodes of type campaign, phone number, or URL when selecting meta-paths between two users. 

%% file: method.tex
\section{Proposed Methodology}\label{sec:method}
\begin{figure}[h]
\includegraphics[width=0.95\linewidth]{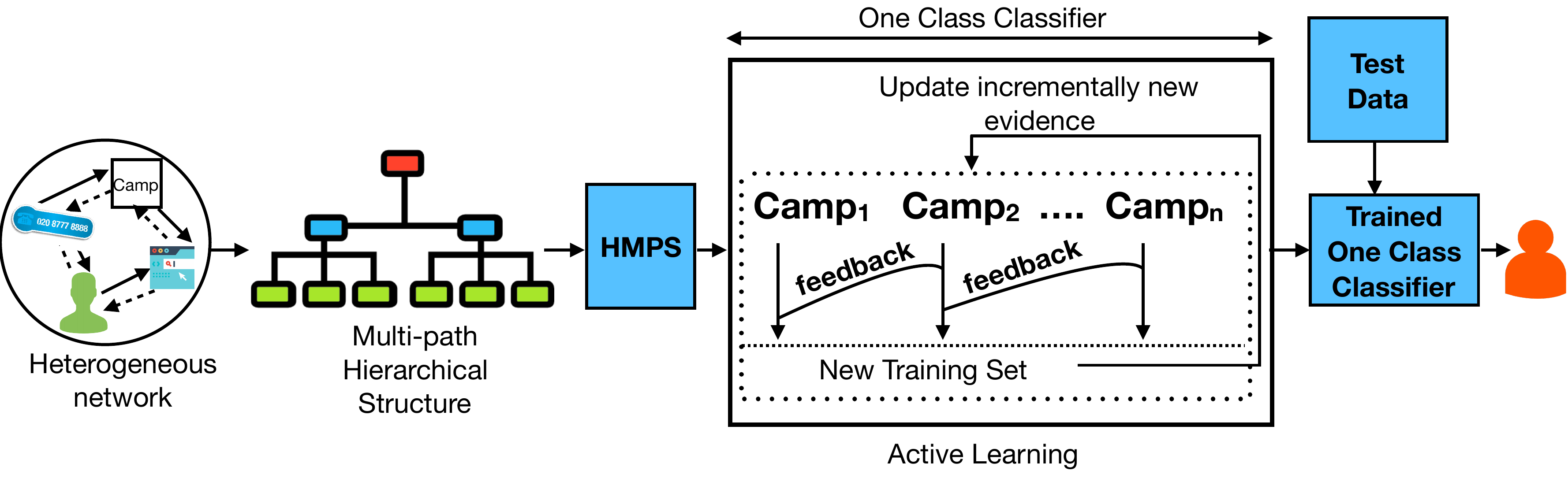}
\caption{Proposed collective classification framework to detect spammers on Twitter.}
\label{fig:system_arch}
\vspace{-5mm}
\end{figure}

In this section, we describe the overall proposed methodology for collectively classifying users as spammers on Twitter (see Figure~\ref{fig:system_arch}). 

\textbf{Why collective classification?} Collective classification refers to the combined classification of nodes based on correlations between unknown and known labels~\cite{sen2008collective}.
Given the labels of the instances in training set $Tr \subset All$,  the task of collective classification in HIN is to infer the labels of the testing set ($Te = All - Tr$). We address collective classification problem using HMPS to find users (unknown labels) that are \emph{similar} to spammers (known labels).
In individual classification,  nodes are classified individually without taking into account their interdependencies via the underlying network structure. However, in our heterogeneous networks, nodes are connected by same phone number or URL. Therefore, we employ collective classification approach. It has been shown to achieve better accuracy compared to independent classification~\cite{sen2008collective}.

\subsection{Hierarchical Meta-Path Scores (HMPS) \label{hmps}}
After identifying campaigns~(see Section~\ref{campaignidentification}), the next step is to measure HMPS for a user (Algorithm~\ref{alg:hmpsalgo}) to find the similarity of the user with other known spammers (suspended users). To this end, we propose an additive similarity score for a user with respect to all the spammers in that particular campaign. 
Although there are several other similarity measures available, they are biased towards underlying network structure and prior information about relevant meta-paths. For instance, PathSim~\cite{sun2011pathsim} only works for symmetric relations, HeteSim~\cite{shi2014hetesim} relies on the relevance of a single meta-path. Forward Stagewise Path Generation (FSPG)~\cite{meng2015discovering} generates the set of most relevant meta-paths under a given regression model, which is validated by a human expert. However, in the context of Twitter being modeled as a HIN, the relevant meta-paths are not known. Therefore, it is computationally intractable to find the relevance of a meta-path.

This motivates us to propose a novel meta-path based similarity measure, called {\em Hierarchical Meta-Path Scores} ({\em HMPS}) that captures the similarity between two users based on the function of distance through which they can be reached. 

\begin{figure}[h]
\begin{center}
\includegraphics[width=0.65\columnwidth]{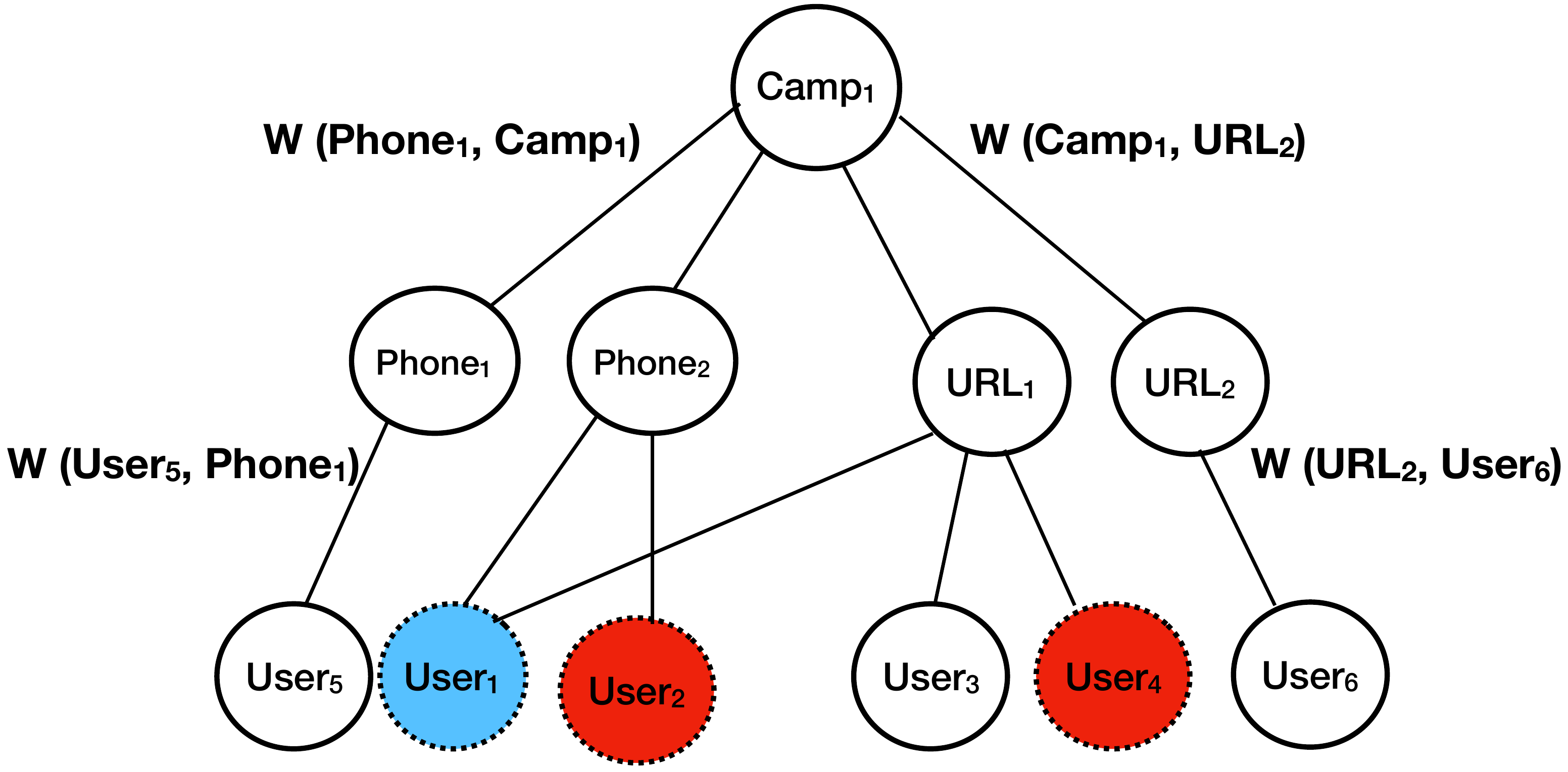}
\caption{A hierarchal structure to measure HMPS of users. Users with red color are known spammers.} 
\label{fig:dependencetree}
\end{center}
\end{figure}

\textbf{HIN to hierarchical structure:} To measure HMPS, we model the Twitter heterogeneous network in the form of a multi-path hierarchical structure as shown in Figure~\ref{fig:dependencetree}. In this structure, nodes on a meta-path are connected with their Least Common Ancestor~(LCA) node. LCA node for users is taken as a phone number or URL, and subsequently, campaign node is taken as the LCA node for a phone number / URL. The purpose of LCA node is to limit the range of operations that can be applied across two related nodes. We choose such a structure because if two users share the same phone number or URL for promoting campaigns, they should be more similar rather than two users who do not share any common phone number or URL but are still part of a single campaign. 
The intuition behind HMPS is that if two users are strongly connected to each other, the distance between them in the hierarchical structure would be less. 

The \emph{similarity score} between two entities \emph{x} and \emph{y} is a real number, computed by a function $F$ of the similarity scores for each meta-path of a set $\Theta :\Psi (x,y \mid \Theta) = F(max\{\phi (x,y \mid \Pi_{i}) \mid 1 \leqslant i \leqslant p)\})$, where $\phi(x,y \mid \Pi_{i})$
is a similarity score between $x$ and $y$ given meta-path $\Pi_{i}, \Theta = {\Pi_{1},...,\Pi_{p}}$, and $F$
is the maximum similarity score over the `p' meta-paths. Then the HMPS of an entity $x$ is defined as: $HMPS(x)=\sum_{y\in S} \Psi(x,y)$, where $S$ is the set of spammers in the campaign where $x$ belongs to.  

For every user, HMPS is calculated with respect to each spammer (suspended user) in the campaign, and the scores are finally added, as shown in Algorithm~\ref{alg:hmpsalgo}.
Following are the weights used for each edge in the hierarchical structure.
\begin{itemize}
\item \textbf{W($User_{i}$, $Phone_{j}$):} This is the weight of the edge connecting a user and a phone number, and is measured as the ratio of tweets propagated by $User_{i}$ containing $Phone_{j}$ over all the tweets containing $Phone_{j}$.

\item \textbf{W($User_{i}, URL_{j}$):} This is the weight of the edge connecting a user and a URL, and is measured as the ratio of tweets propagated by $User_{i}$ containing $URL_{j}$ over all the tweets containing $URL{j}$.

\item \textbf{W($Camp_{i}, Phone_{j}$):} This is the weight of the edge connecting a campaign and a phone number, and is measured as the ratio of tweets containing $Phone_{j}$ in $Camp_{i}$ over cumulative frequency of URLs and phone numbers in $Camp_{i}$.

\item \textbf{W($Camp_{i}, URL_{j}$):} This is the weight of the edge connecting a campaign and a URL, and is measured as the ratio of tweets containing  $URL_{j}$ in $Camp_{i}$ over cumulative frequency of URLs and phone numbers in $Camp_{i}$.
\end{itemize}

Let us assume that we want to calculate the HMPS for $User_{1}$ (unknown) shown in Figure~\ref{fig:dependencetree}. The campaign contains two suspended users, $User_{2}$ and $User_{4}$. So the HMPS score of $User_{1}$ is calculated w.r.t. $User_{2}$ and $User_{4}$ as follows:
\begin{itemize}

\item Weight between $User_{1}$ and $User_{2}$, $W_{1}$: $W~(User_{1},Phone_{2})$ . $W~(User_{2},Phone_{2})$

\item Weight between $User_{1}$ and $User_{4}$, $W_{2}$: maximum score calculated for 2 possible meta-paths, i.e., \texttt{User\textsubscript{1}-URL\textsubscript{1}-User\textsubscript{4}} and \texttt{User\textsubscript{1}-Phno\textsubscript{2}-Camp\textsubscript{1}-URL\textsubscript{1}-User\textsubscript{4}}; $W_{2}$ = max ({W~$(User_{1},URL_{1})$ .$W~(User_{4},URL_{1})$}, {$W~(User_{1},Phone_{2})$ . $W~(Camp_{1},Phone_{2})$ . $W~(Camp_{1},URL_{1})$ . $W~(User_{4},URL_{1})$}) 

\item The final HMPS of $User_1$, HMPS~($User_{1}$)= $W_{1}$ + $W_{2}$.
\end{itemize}

\begin{algorithm}[h]\small
\caption{HMPS for Collective Classification}
\begin{algorithmic}[1]
\For{$Camp_i\in Campaigns$}

\State $S$ = Set of known spammers in $Camp_i$ ($m=|S|$);  $U$ = Set of unknown users in $Camp_i$; $n$ = Total number of users in $Camp_i$ 

\State $ score_{i} \gets  \sum_{j=1}^{m} \text{HMPS} (U_{i}, S_{j}, Camp_i) \forall i \in [1, n] $

\EndFor

\Procedure {HMPS} {$u$, $s$, $camp$}
\State $res = 0$ 
\For{$i \in  Parent(u)$} \Comment{$Parent(u)=$ Immediate antecedent of $u$}
\For{$j\in Parent(u)$}
\If {$i == j$}
\Comment{$W(s,j)$=weight of the edge $\langle s,j \rangle$ in the hierarchical structure}
\If {$W(u,i) . W(s,j) > res$} \State $res \gets W(u,i). W(s,j)$ 
\EndIf 
\Else
\If{$W(u,i) . W(s,j) .  W(i,camp) . W(j,camp) > res$}
\State $res \gets W(u,i) . W(s,j) . W(i,camp) . W(j,camp)$ 
\EndIf
\EndIf
\EndFor
\EndFor
\State \textbf{return} $res$
\EndProcedure
\end{algorithmic}
\label{alg:hmpsalgo}
\end{algorithm}

Note that in order to measure the HMPS for each user from the hierarchical structure, we build the hierarchical structure for individual campaigns separately instead of combining all the campaigns due to the following two reasons: (i) it is computationally expensive to find meta-paths for all the connections of users across campaigns from a large hierarchical structure, and (ii) HMPS is an absolute value; global HMPS can result in wrong labeling. Specifically, if a spammer~($S$) has HMPS value $X$ in campaign $C_{1}$ and other unknown user~($U$) has same value $X$ in another campaign $C_{2}$, then $U$ will be wrongly labeled as a spammer. It might not be a spammer based on HMPS calculated within that campaign.

\subsection{Active Learning with Feedback \label{active learning}}
As we consider only those campaigns which contain more than one suspended user (spammer), the classes (spammers and non-spammers) present in our dataset would be highly imbalanced. 
Exiting research has shown that \textbf{\textit{one-class classification}} \textbf{(\textit{OCC})} achieves much better performance than two-class classification if: (i) there is highly imbalanced dataset~\cite{raskutti2004extreme} and the target class is prevalent in the training set, (ii) the unknown instances do not belong to any known class, or (iii) the unknown instances are difficult to be categorized into a known class due to several reasons such as lack of annotators, lack of enough evidences etc.  
OCC is trained only on the target class (which is spam in our case), and its task is to define a classification boundary around the target class, such that it accepts as many instances as possible from the target class, while it minimizes the chance of accepting the outlier instances. In OCC, since only one side of the boundary can be determined, it is hard to decide from just one-class how tightly the boundary should fit in each of the directions around the data. It is also hard to decide which features should be used to find the best separation of the target and outlier class instances. 

\begin{figure}[h]
\begin{center}
\subfigure[Number of suspended users per campaign.]
{ \label{fig:campaign_sus}
\includegraphics[width=0.45\linewidth]{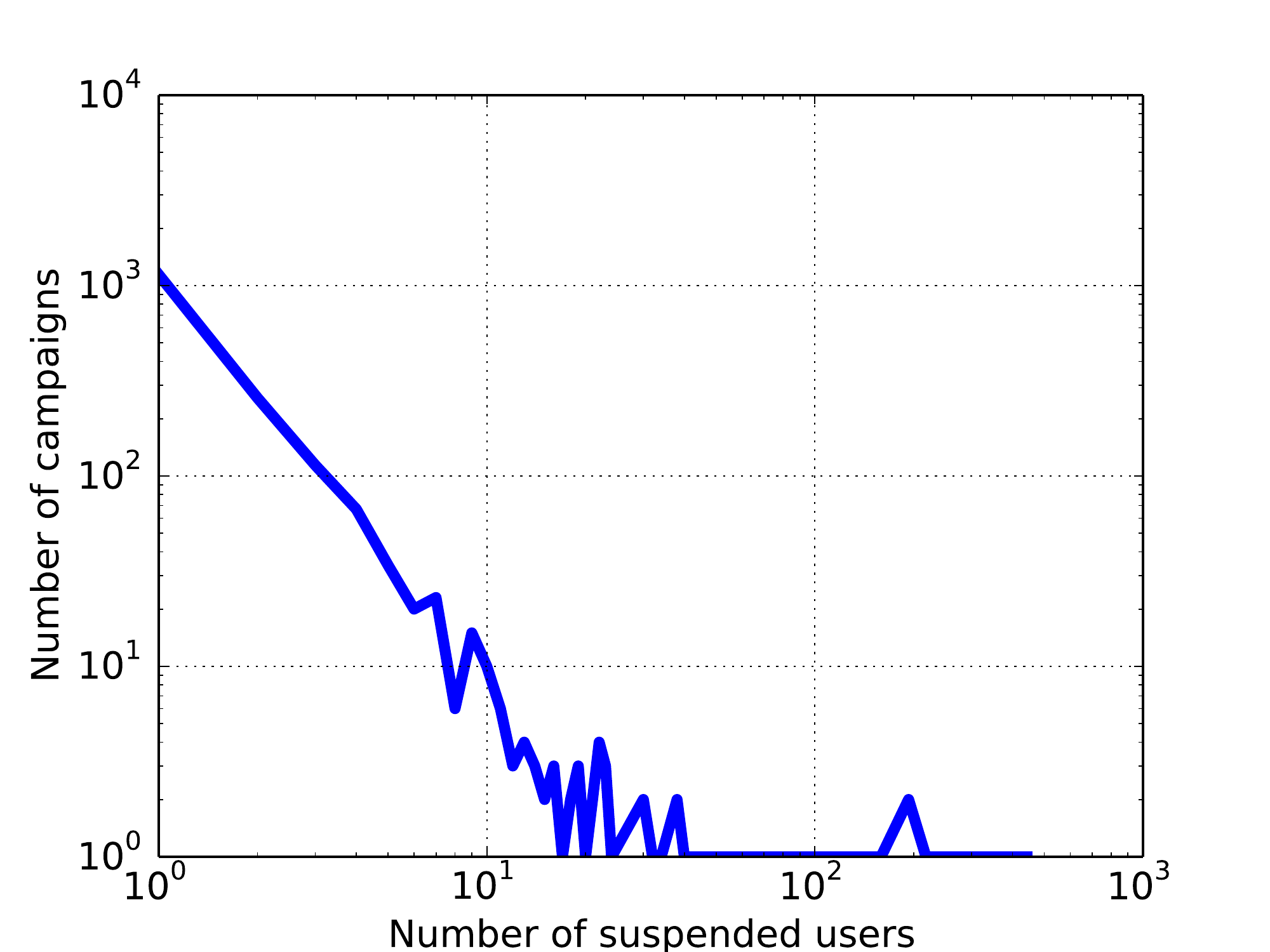}
}
\subfigure[Overlapping users in campaings.]
{ \label{fig:usersoverlap}
\includegraphics[width=0.45\linewidth]{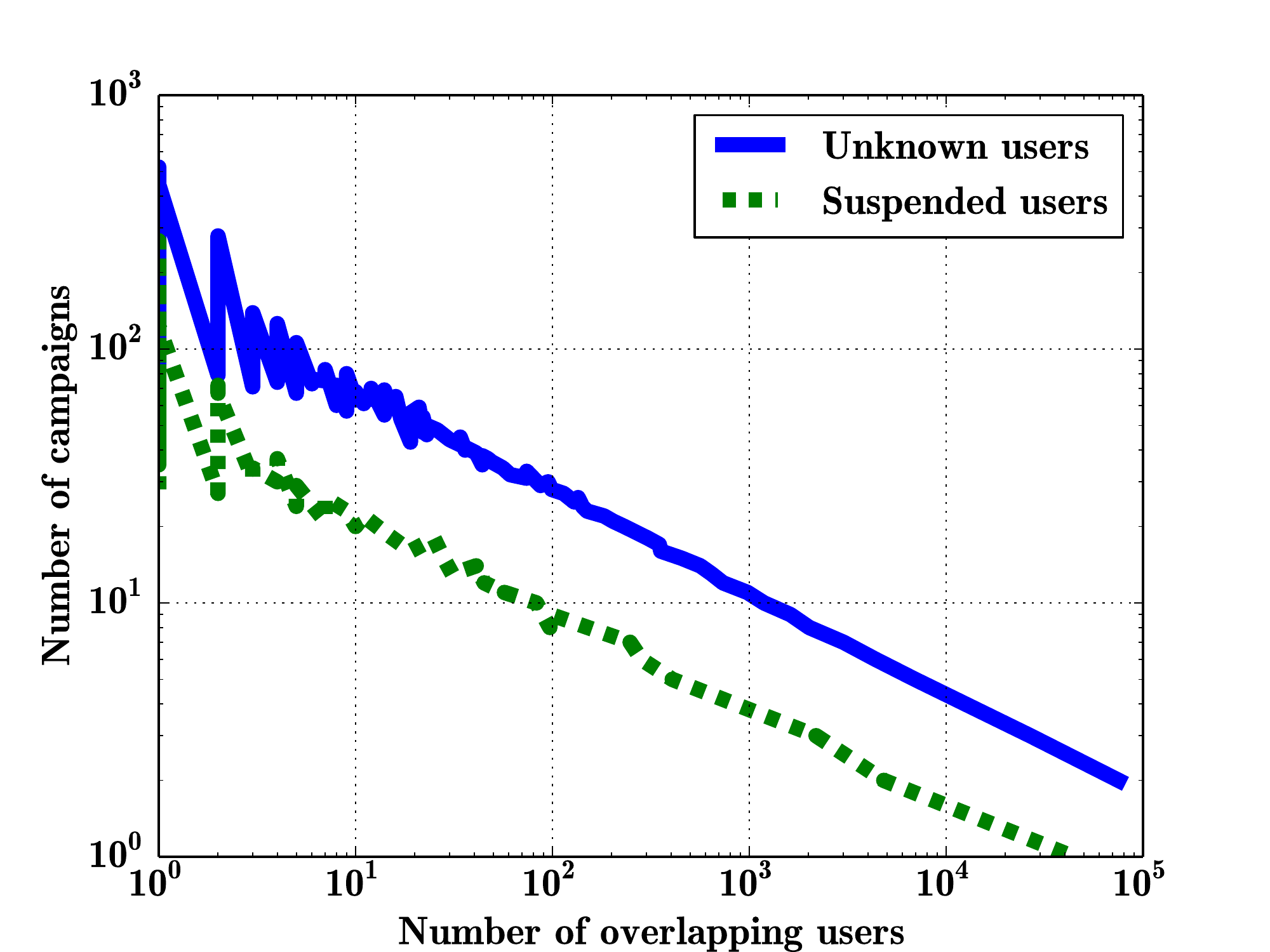}
}
\caption{Distribution of the (a) suspended and (b) overlapping users (users belonging to multiple campaigns) in our dataset. The number of suspended users per campaign is less. Therefore, to increase the training samples, overlapping users are picked for human annotation.}
\label{fig:allaboutusers}
\vspace{-5mm}
\end{center}
\end{figure}

\textbf{Learning with feedback:} We would like to reiterate that we picked individual campaigns and not the entire dataset together since the HMPS local to a campaign helps in finding similar users better~(see Section~\ref{hmps}). Each campaign is associated with a supervised classifier (one-class classifier in our case). Out of $3,370$ campaigns in the dataset that have at least one suspended user, not all campaigns have sufficient training samples to train the models, as shown in Figure~\ref{fig:campaign_sus}. 
However, the process of human annotation to enrich the training set can be costly. 
To reduce the effort of human labeling, one can obtain meaningful shreds of evidence from some external sources and incorporate them into the training set. For instance, in ensemble learning, one can leverage the output class of unknown objects obtained from one classifier and feed them into the other classifiers. This might be related to \emph{active learning}, where given a pool of unlabeled data, one can try to select a set of training examples actively to reach a minimum classification error.

Since individual campaigns may not have significant training instances, we propose an {\em active learning approach with feedback} to collect cues about unknown users from multiple campaigns to enlarge the individual training set associated with each campaign-specific model. We further notice that campaigns have significant user overlap --  21\% users belong to multiple campaigns (see Figure~\ref{fig:usersoverlap} for the distribution of overlapping users). Presence of user overlap further motivates us to incorporate the feedback-based model as follows. 

\begin{figure}[h]
\includegraphics[width=0.75\linewidth]{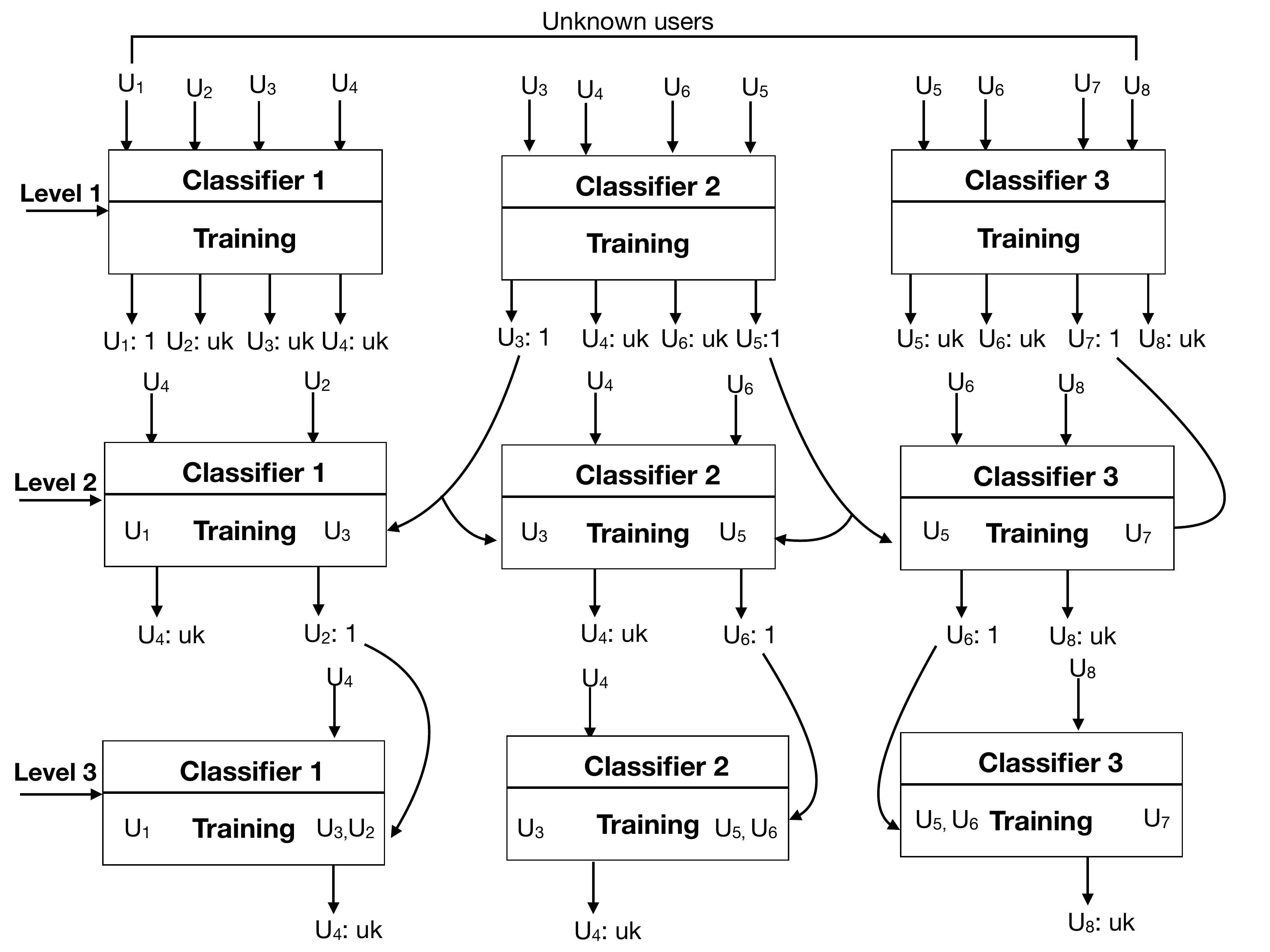}
\caption{A schematic diagram of active learning with feedback amongst campaign-specific classifiers.}
\label{fig:activelearning}
\vspace{-5mm}
\end{figure}

Let us assume that user $u$ is classified as a spammer by a classifier (associated with a campaign say, $Cam_i$) with high confidence. If $u$ is also a part of some other campaigns (say, $Cam_j$) where the class of $u$ is unknown, we assign $u$ to the training set of $Cam_j$ along with its class as a spammer. In this way, we keep increasing the size of the training set of individual classifiers (see Figure \ref{fig:activelearning} for a schematic diagram of our proposed feedback-based active learning method).
Overall, we perform the following steps:
\begin{itemize}
\item An initial set of labeled instances is used for training individual classifiers. Since one-class classifier is used, the training set consists of only the spammers (suspended Twitter accounts). Each campaign-specific classifier is then used to label the unknown users. 
\item From each set of unknown users labeled by the classifier, we choose a subset of users according to the \emph{selection criterion} (mentioned later). The selected users are then augmented with the training set of other classifiers whose corresponding campaigns also contain these users. 
\item These steps are iteratively executed for all the campaigns. This constitutes level 1 of the iteration (as shown in Figure \ref{fig:activelearning}). At the end of this level, we obtain a set of new training set for each classifier.
\item In the next level, the new training set is introduced to the classifier and used to predict the class of the rest of the unknown users. This constitutes level 2 of the iteration. The above process convergences once we obtain no more labeled user from the current level to be augmented further with the training set of any classifier in the next level.
\end{itemize}

\textbf{Selection criterion:} It is important to decide a selection criterion to choose a subset of users from the output of the classifiers; inappropriate criterion might inject noise in the training set that will propagate throughout succeeding levels. We propose the following criterion for selecting users:
\begin{displayquote}
Given (a) a one-class classifier $C$, represented by the function $f(x)$ which, for
an instance $x$, provides the distance of $x$ from the classification boundary, and (b) $X$, a set of unlabeled instances, 
we take the maximum distance among all the training samples from the decision boundary, $T_{max}^c=\max_{x\in X} f(x)$. Now, from the unknown set $X_u$ which are labeled by $C$, we choose those instances $X'_{u}$ such that $\forall x\in X'_{u}: f(x)\geq T_{max}^c$. Note that the threshold $T_{max}^c$ is  specific to a campaign.
\end{displayquote}

%% file: results.tex
\section{Experimental Results}
In this section, we start by presenting the baseline methods used to compare with our method, followed by a detailed comparative evaluation.

\subsection{Baseline Methods} \label{baseline}
We compare our method with three state-of-art methods proposed in the literature for spam detection in general. However, none of them focused on phone number specific spammers whose dynamics are different (as mentioned in Section \ref{dataset}). Since we did not obtain the source code, we implemented the methods on our own. Note that all the baselines originally used 2-class classifiers. However, in this paper, we show the results of the baselines both for one-class and 2-class classifications after suitable hyper-parameter optimization. 

\textbf{Baseline 1:} We consider the spam detection method proposed by Benevenuto et al. ~\cite{benevenuto2010detecting} as our first baseline. They proposed the following OSN-based features (referred as {\bf OSN1}) per user: fraction of tweets with URLs, age of the user account, average number of URLs per tweet, fraction of followers per followee, fraction of tweets the user replied, number of tweets the user replied, number of tweets the user receives a reply, number of friends and followers, average number of hashtags per tweet. They showed that the SVM-based classifier performs best. 

\textbf{Baseline 2:} We consider the method proposed by Khan et al. \cite{khan2016segregating} to segregate spammers from genuine experts on Twitter as our second baseline. They suggested the following features (referred as {\bf OSN2}): authority and hub scores of users in the follower-followee network, fraction of the user's tweets that contain the URLs, average number of URLs in a tweet, average number of URLs per number of words in a tweet of the user, average number of hashtags per number of words in a tweet, and average number of hashtags in a tweet. They showed that Logistic Regression performs best. 

\textbf{Baseline 3:} We consider the method proposed by Adewole et al. ~\cite{adewole2017smsad} to detect spam messages and spam user accounts as our third baseline. They proposed the following list of profile and content-based features (referred as {\bf OSN3}): length of the screen name based on characters, the presence or absence of profile location, whether the user includes URL or not in his profile, age of the account in days,  number of followers of the user, number of friends~/~followers of the user, total statuses of the account, number of tweets the user has favorited, indicating presence or absence of profile description, whether the user has not modified the theme of their profile, presence or absence of time zone, whether the account has been verified or not, whether the user has not changed the default profile egg avatar, number of the public lists the user is a member, whether or not the user has enabled the possibility of geo-tagging their tweets, normalized ratio of followers to friends, ratio of the number of follower to friends, ratio of the number of friends to followers, (total, unique, and mean) number of tweets, hashtags, URLs, mentions, favorite count, and retweets, ratio of (hashtags, URLs,  mentions, retweets) to total number of tweets, (hashtag, URLs, mention, retweet, tweet-length) deviation, average number of daily tweets, average tweet length, popularity ration, number of duplicate tweets, and maximum value of hashtag frequency. They showed that Random Forest performs best for the classification task. 

Note that previous work considered only those campaigns which involve only URLs~\cite{benevenuto2010detecting,khan2016segregating, adewole2017smsad}. In our work, a phone number, being a stable resource, helped in forming campaigns better. Besides, most of the OSN features used in the baselines are easy to evade by spammers, whereas HMPS-based feature is difficult to manipulate.

\subsection{Experimental Setup}\label{setup}
Our proposed classification method is run separately with different features (HMPS, OSN1, OSN2, and OSN3) and their combinations. We use the standard grid-search technique to tune the hyper-parameters. For evaluation, we design two experimental settings: 

\noindent {\bf (i) Setting 1:} Our primary goal is to detect user accounts which are suspended by Twitter because they are spam accounts. Therefore, the set of suspended accounts constitutes the ground-truth for the classifiers. 
Out of all suspended accounts present in our dataset (mentioned in Section \ref{dataset}), we adopt leave-one-out cross-validation technique (due to the very limited number of suspended accounts) and report the average accuracy of the classifiers. Note that in this setting, we use one-class classifier for all the competing methods. 

\noindent {\bf (ii) Setting 2:} We believe that our method is capable of detecting those accounts which are spammers, but not suspended by Twitter yet. Therefore, we further invited {\bf human annotators}\footnote{All annotators were security researchers between the age group of 25~-~35 years.} to annotate some non-suspended accounts as spammers or non-spammers. This will further help us to run the baseline methods which originally used binary classifiers (see Section \ref{oneclasstwoclass}). Since it is not possible to label all non-suspended users, we adopt a convenient sampling approach. We define user bins according to the number of campaigns the non-suspended users exist (see the distribution in Figure \ref{fig:usersoverlap}). Our sampling approach preferentially chooses users who are part of multiple campaigns to maximize the evidence per campaign --  the probability of choosing a user belonging to multiple campaigns is higher than that for a user who is a part of a single campaign. Following this approach, we picked $700$ users from $3,370$ campaigns. Each user was labeled by three human annotators as spammers or non-spammers, and then the majority vote was considered as the final class. The inter-annotator agreement was $0.82$ according to Cohen's kappa measure.

Out of 700 manually annotated accounts, we hold out 20\% of the dataset to be used as the test set in Setting 2. We repeat this experiment 50 times and report the average accuracy. Here also, we use one-class classifier for all the competing methods and consider `spammer' as our target class.

{\bf Evaluation metrics:} For comparative evaluation, we use the standard information retrieval metrics -- Precision, Recall, F1-score, Area under the ROC curve (AUC).

\subsection{Comparative Evaluation} \label{baseline}
Table \ref{result:setting2} shows the performance of the competing methods for both settings. We report the results of our active-learning based one-class classifier with different feature combinations.~\footnote{We tried with other combinations as well such as HMPS+OSN1+OSN2, HMPS+OSN2+OSN3 etc. The results were not encouraging enough to be reported in the paper.} For setting 1 (leave-one-out), we report the performance w.r.t the {\em accuracy} (fraction of known spammers identified by the method) and observe that our method performs significantly well with only HMPS feature -- it achieves an accuracy of $0.77$, outperforming all baseline methods. However, incorporating OSN2 features along with HMPS further enhances 9.1\% performance of our classifier, achieving an accuracy of $0.84$. 

A similar pattern is observed for setting 2. However, here our model with only HMPS turns out to be even stronger classifier, outperforming all others in terms of precision ($0.99$), F1-score ($0.93$) and AUC ($0.88$). Here also, incorporating most of the OSN features with HMPS does not enhance the performance of our method (or sometimes deteriorates the performance), except ONS2 which seems to be quite competitive. However, baseline 2 seems to be the best method w.r.t recall ($0.92$); but it significantly sacrifices the performance w.r.t. precision, F1-score, and AUC.

Nevertheless, {\bf we consider the following setting as our default method since it outperforms other methods in almost all experimental setup: HMPS + OSN2 + one-class classifier + active learning}.  Baseline 2 is considered as the best baseline method in the rest of the paper. 

\begin{table}[h]
\centering
\caption{Comparative evaluation of the competing methods on two different experimental settings. For all the methods, one-class classifier is used. The colored row shows the performance (P: Precision, R: Recall, F1: F1-score) of our default method. The last row shows the results of our default method {\em without} active learning (see Section \ref{feedbackresults}). }
\label{result:setting2}
\scalebox{0.85}{
\begin{tabular}{|c|c|c|c|c|c|c|c|}
\hline 
\multirow{2}{*}{\textbf{Method}} & \multirow{2}{*}{\textbf{Feature}} & {\bf Setting 1} & \multicolumn{4}{|c|}{{\bf Setting 2}}\\\cline{3-7}
  &  & {\bf Accuracy} & \textbf{P} & \textbf{R} & \textbf{F1} & \textbf{AUC} \\ \hline \hline
Baseline 1 & OSN1       &  0.62           & 0.86               & 0.71           & 0.77         & 0.48             \\ \hline
Baseline 2 & OSN2        &     0.58          & 0.84               & {\bf 0.92}           & 0.87         & 0.52             \\ \hline
Baseline 3 & OSN3        &         0.62     & 0.86               & 0.66           & 0.74         & 0.47             \\ \hline
\multirow{4}{*}{Our} & HMPS               &             0.77 & {\bf 0.99}               & 0.87           & {\bf 0.93}         & {\bf 0.88}             \\ \cline{2-7}

 & HMPS + OSN1        &          0.76  & 0.89               & 0.90           & 0.89         & 0.72             \\ \cline{2-7}
 & {\color{blue}HMPS + OSN2}              &    {\color{blue} {\bf 0.84}}   & {\color{blue}0.98}               & {\color{blue}0.88}           & {\color{blue}{\bf 0.93}}         & {\color{blue}0.87}             \\ \cline{2-7}
 & HMPS + OSN3 &            0.70     & 0.88               & 0.73           & 0.80          & 0.59             \\ \hline\hline

\multirow{2}{*}{Our} & HMPS + OSN2 & \multirow{2}{*}{--} &\multirow{2}{*}{0.42} & \multirow{2}{*}{0.98} & \multirow{2}{*}{0.55} & \multirow{2}{*}{0.51}\\
 & - Active Learning & & & & &\\\hline
\end{tabular}}
\vspace{-3mm}
\end{table}

\noindent {\bf Justification behind superior performance of HMPS:} \label{hmpsgood}
All of the baseline methods rely on the features that can be changed over time. These methods either consider URL attributes (baselines 1 and 3) within the tweets or changes in profile characteristics between a legitimate and spam user account (baselines 2 and 3). 
Given these specificities, it is easy for a spammer to manipulate these features. In contrast, HMPS relies on the monetization infrastructure~(phone numbers) to identify campaigns and spammers. As discussed earlier, we aggregate tweets as part of the same campaign when they use multiple phone numbers wrapped around similar text. As a result, our method is resilient to spammers' manipulation. 
Furthermore, to understand how HMPS helps in improving the detection of spammers over the baselines, 
we manually analyze a sample of `spammers'. Some of the users not identified by baselines 1 and 3 as spammers have a balanced number of friends and followers and a low number of tweets. In addition, users were not using URLs to spread the campaign. Therefore, all URL-based features do not aid in the detection task. 

Baseline 2 measures the authority and hub scores based on the tweets with hashtags. As a result, it wrongly detects some benign
users as spammers that were retweeting posts related to (say,) blood donation campaigns. When baseline 2 is combined with HMPS, the false positive rate is reduced since these users are not found in the spammer network.

\begin{figure}[h]
\vspace{-4mm}
\begin{center}
\subfigure[]
{ \label{fig:spammerbio}
\includegraphics[width=0.23\linewidth]{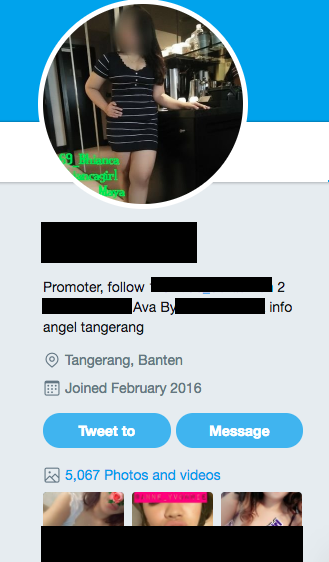}
}
\hspace{1cm}
\subfigure[]
{ \label{fig:spammertweets}
\includegraphics[width=0.30\linewidth]{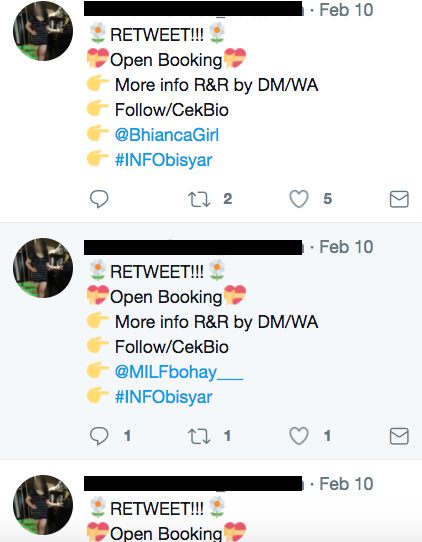}
}
\caption{An example spammer account (bio shown in (a), timeline shown in  (b)) that has not been suspended by Twitter yet, but our system could detect it as spammer.}
\label{fig:twitternotsuspended}
\vspace{-5mm}
\end{center}
\end{figure}

In addition, HMPS can find spammers that are not suspended by Twitter yet. For instance, Figure~\ref{fig:twitternotsuspended} shows a spammer account that clearly violates the Twitter policy by promoting and posting repeated, pornographic content.~\footnote{https://support.twitter.com/articles/18311} Surprisingly, this account has not been suspended by Twitter yet. However, we found similar such accounts suspended by Twitter. Interestingly, our system was able to identify this account as a spammer. 

These examples show that HMPS can identify spammers that use phone numbers, which are not detected by the baseline systems and~/~or Twitter, and is, therefore, more effective in detecting spammers that spread phone numbers to promote campaigns.

\subsection{One-class vs. 2-class Classifier} \label{oneclasstwoclass}
One may argue that the results reported in Table~\ref{result:setting2} may not reflect the original performance of the baseline methods since all the baseline methods originally used 2-class classifiers. Moreover, there was no empirical justification for adopting one-class classifier over 2-class classifier. To address these arguments, here we exactly replicate the baseline methods by considering the best 2-class classifier per baseline reported in the papers. We choose a balanced dataset of $150$ suspended and 150 non-suspended users randomly sampled from our manually labeled dataset (see setting 2 in Section \ref{setup}). For comparative evaluation, we consider several state-of-the-art 2-class classifiers (Logistic regression (LR), Latent Dirichlet Allocation (LDA), K-nearest neighbors (KNN), Decision Tree (DT), Naive Bayes (NB), Random Forest (RF), and Support Vector Machine (SVM)) and adopt them into our active learning framework. 
Table~\ref{onevstwo} shows that none of the baselines and our adopted 2-class classifiers outperform our default one-class classifier (last row of Table \ref{onevstwo}). Our default method is 12.6\%, 7.7\%, 10.7\% and 9.7\% higher than the second-ranked method (Decision Tree) in terms of precision, recall, F1-score, and AUC respectively. This result indicates that one-class classification is always helpful for the application where there is limited labeled data, and the label of most of the instances is unknown.

\begin{table}[h]
\centering
\caption{Results of 2-class classifiers and comparison with our default one-class classifier. Here, the best 2-class classifiers reported in the papers are considered for the baselines.}
\label{onevstwo}
\vspace{-3mm}
\scalebox{0.8}{
\begin{tabular}{|c|c|c|c|c|}
\hline
\textbf{Method}        & \textbf{Precision} & \textbf{Recall} & \textbf{F1-score} & {\bf AUC} \\ \hline
Baseline 1 & 0.68 &0.69 & 0.65 & 0.50\\\hline
Baseline 2 & 0.47 &0.57 & 0.51 & 0.50\\\hline
Baseline 3 & 0.79 &0.78 & 0.78 &  0.57\\\hline\hline
\multicolumn{5}{|c|}{{\bf HMPC + 2-class classifier}}\\\hline
LR            & 0.61              & 0.58            & 0.55       &       0.58 \\ \hline
LDA           & 0.61              & 0.58           & 0.55         &     0.58 \\ \hline
KNN           & 0.75             & 0.74            & 0.74          & 0.74   \\ \hline
DT & 0.83              & 0.83           & 0.83         &   0.83  \\ \hline
NB   & 0.60              & 0.58            & 0.57          &    0.58\\ \hline
SVM           & 0.65              & 0.63            & 0.62 &        0.63     \\ \hline
RF & 0.83              & 0.82            & 0.82              & 0.82 \\ \hline\hline
\multicolumn{5}{|c|}{{\bf Our default one-class classifier}}\\\hline
HMPS+OSN2          & \textbf{0.95}              & \textbf{0.90}            & \textbf{0.93}   & \textbf{0.92}           \\ \hline
\end{tabular}}
\vspace{-4mm}
\end{table}

\subsection{General vs. Active Leaning} \label{feedbackresults}
We argue that since the number of training samples is not sufficient for individual classifiers, feedback from one classifier to another would increase the training set that in turn enhances the performance of the classifier. To verify our argument, we run our default method without active learning (without feedback) and observe that although recall increases significantly (0.98), it degrades the performance w.r.t. other performance measures --  57.1\%, 40.8\% and 41.3\% degradation of precision, F1-score and AUC respectively (see the last row of Table \ref{result:setting2}). High recall with low precision indicates that most of the unknown users are classified as spammers by the general classifier. It happens due to the limited labeled data, which active learning can efficiently handle.

\subsection{Feedback vs. Oversampling} \label{feedbackoversample1}
Since the size of the training set is small for each campaign-specific classifier, we use feedback across campaigns to increase the training set. As an alternative, one can also use other state-of-the-art oversampling techniques such as SMOTE~\cite{chawla2002smote}. Here, we adapt SMOTE for increasing the size of the training data~(i.e., target class: spammers). The training set is oversampled by taking each training sample and introducing synthetic examples along the line segments joining any~/~all $k$ neighbors of the training sample. Depending on the amount of oversampling required, $k$-nearest neighbors are randomly chosen. Table~\ref{feedbackoversample} shows the results for different values of oversampling ratio, i.e., the fraction of training set taken as the number of synthetic samples.
In addition, we perform the oversampling technique before dividing the data into training and validation to ensure that the information from the training set is used in building the classifier. Table~\ref{feedbackoversample} shows that even after varying the ratio for oversampling, none of the cases can achieve the accuracy obtained from our feedback-based learning approach. This indicates that our feedback-based learning strategy is superior to the other oversampling strategy.

\begin{table}
\centering
\caption{Comparison of our feedback-based learning approach with standard oversampling approach (SMOTE). The term `Ratio' indicates the fraction of training set taken as the number of synthetic samples generated by the oversampling technique.}
\label{feedbackoversample}
\vspace{-3mm}
\scalebox{0.9}{
\begin{tabular}{|l|c|c|c|c|}
\hline
\multicolumn{5}{|c|}{{\bf Oversampling + default one-class classifier}}\\\hline
 & Precision & Recall & F1-Score & AUC \\\hline
Ratio = 0.20          & 0.90              & 0.64            & 0.64   & 0.59           \\ \hline
Ratio =  0.30         & 0.88              & 0.74            & 0.74   & 0.63           \\ \hline
Ratio = 0.50          & 0.81             & 0.71            & 0.68   & 0.58           \\ \hline
Ratio = 0.75          & 0.91              & 0.68            & 0.69   & 0.56           \\ \hline
Ratio = 1          & 0.91              & 0.68            & 0.70   & 0.57           \\ \hline
\multicolumn{5}{|c|}{{\bf Feedback + default one-class classifier}}\\\hline
& \textbf{0.95} & \textbf{0.90} & \textbf{0.93} & \textbf{0.92}\\\hline
\end{tabular}}
\vspace{-3mm}
\end{table}

%% file: related.tex
\section{Related Work}
In this section, we highlight the prior work on classification, clustering, and similarity tasks carried out using meta-paths. We also discuss several existing techniques to identify spammers.

\textbf{Meta-path classification:} Sun et al.~\cite{sun2011pathsim} first proposed the idea of meta-path in heterogeneous network. Since then, it  has been used extensively in various applications such as  classification~\cite{kong2013multi,li2016transductive}, clustering~\cite{sun2013pathselclus}, and similarity measures~\cite{shi2012relevance,sun2011pathsim}. Sun et al. proposed a measure called ``PathSim'' which outperformed Path Constrained Random Walk~(PCRW) proposed by ~\cite{lao2010relational}.
Meng et al. introduced biased constraint random walk to handle both symmetric and non-symmetric meta-paths~\cite{meng2015discovering}.
They proposed Forward Stagewise Path Generation algorithm (or FSPG), which derives meta-paths that best predict the similarity between a node pair.
Shi et al. proposed ``HeteSim'' to measure the relevance of any node pair in a meta-path~\cite{shi2014hetesim}. 
To overcome the computational and memory complexity of HeteSim, Meng et al. proposed ``AvgSim'' that measures similarity score through two random walk processes along the given meta-path and the reverse meta-path~\cite{meng2014relevance}. 
Besides these similarity measures, Zhang et al. found node similarity based on connections between centers in X-star network~\cite{zhang2015top}.
While previous work considered meta-paths for classification assuming relevant meta-paths are known or finding greedy approaches to identify relevant meta-paths, in this work, we model Twitter as a heterogeneous network where relevant meta-paths are unknown. We further propose Hierarchical Meta-Path Scores to predict an unknown user as a spammer based on its neighboring spammers.

\textbf{Spammer classification:}
Previous literature has addressed the problem of spam and spammers on Twitter and other OSNs \cite{thomas2011suspended, gao2010detecting, chu2012detecting, yardi2009detecting, lee2012warningbird, thomas2011design, cao2012aiding, egele2017towards, boshmaf2015integro, cao2014uncovering, danezis2009sybilinfer,liu2015exploiting,stringhini2015evilcohort,ferrara2014detecting}. Benevenuto et al. used OSN based features to detect spammers on YouTube, with an accuracy of 98\%~\cite{benevenuto2009detecting, benevenuto2010detecting}. Researchers have identified spammers on Twitter ~\cite{lee2010uncovering, liu2016detecting, viswanath2014towards} and blogging websites~\cite{khan2016segregating} using user-based features. Previous literature has also looked into identifying fake accounts on OSNs by examining characteristics of user profiles~\cite{stringhini2010detecting}, by learning typical behavior of an account and flagging an account as suspicious in case of deviation~\cite{egele2013compa}, similarity to social bots~\cite{davis2016botornot, ferrara2016rise}, or building a system that considers differences in which legitimate and malicious messages propagate through the network~\cite{nilizadeh2017poised}. Link farming in Twitter where spammers acquire a large number of follower links has been investigated by Ghosh et al.~\cite{ghosh2012understanding}. 
These works treat OSN as a homogeneous network. Modeling OSNs as heterogeneous networks provide us several ways to capture structural similarity between a pair of nodes via meta-paths. In this work, we also demonstrated that our model works better than the baseline models that use only OSN-based features for classification.

%% file: conclusion.tex
\vspace{-2mm}
\section{Conclusion}
In this paper, we detected spammers spreading spam campaigns using phone numbers on Twitter. We modeled Twitter as a heterogeneous network and proposed a collective classification approach that leverages heterogeneous nodes and their interconnections to identify unknown users as spammers. The significant contributions of our method are three-fold: (i) our proposed Hierarchical Meta-Path Score~(HMPS) can efficiently measure how close an unknown user is w.r.t other known spammers; (ii) our proposed feedback-based active learning strategy is effective over three other baselines that use 2-class classifiers for spam detection; (iii) in case of small number of training instances, our proposed feedback strategy performs significantly better than other oversampling strategies. Through a case study and human-annotated dataset, we also showed that our method could find spammer accounts that are not suspended by Twitter yet.